\documentclass[12pt]{article}
\usepackage{graphics}
\usepackage{epsf}
\usepackage{epsfig}
\newcommand{\beq}{ \begin{eqnarray} }
\newcommand{\eeq}{ \end{eqnarray} }
\newcommand{\beqstar}{ \begin{eqnarray*} }
\newcommand{\eeqstar}{ \end{eqnarray*} }

\newcommand{\figscale}{0.35}

\begin{document}
\baselineskip 0.7cm

\begin{titlepage}

\begin{center}

\hfill IPMU 08-0006\\
\hfill KEK-TH-1229\\

{\large 
Inclusive transverse mass analysis for squark and gluino mass determination
}
\vspace{1cm}

{\bf Mihoko M. Nojiri}$^{1}$, 
{\bf Yasuhiro Shimizu}$^{2}$, 
\vskip 0.15in
{\it
$^1${Theory Group, KEK and the Graduate University for Advanced Study (SOUKENDAI), Oho 1-1, Tukuba, Ibaraki, 305-0801, Japan}\\
{Institute for the Physics and Mathematics of the Universe (IPMU),
 University of Tokyo, 5-1-5 Kashiwa-noHa, Kashiwa, Chiba, 277-8568, Japan}\\
}
$^2${Theory Group, Oho 1-1, Tukuba, Ibaraki, 305-0801, Japan}
\vskip 0.15in
and 
\vskip 0.15in
{\bf Shogo Okada}$^{3}$, {\bf Kiyotomo Kawagoe}$^{3}$ 
\vskip 0.15in
$^3${Department of Physics, Kobe University, Kobe 657-8501, Japan}
\vskip 0.5in

\abstract{
We propose an inclusive analysis of a stransverse mass ($m_{T2}$)
using a hemisphere method for supersymmetry studies at the LHC . 
The hemisphere method is  an algorithm to group collinear and high 
$p_T$ particles  and jets, assuming that there are two of such groups 
in a event.
 The $m_{T2}$ is defined as a function of the unknown LSP mass, two 
hemisphere momenta,  and missing transverse momentum. 
The kinematical end point of the $m_{T2}$ distribution 
provides information on the squark and gluino masses.
We perform a Monte Carlo simulation to study the inclusive $m_{T2}$
distribution at the LHC. We show that the end point of the inclusive $m_{T2}$
distribution has a cusp-like structure around the true LSP mass.
The knowledge of the expected kinematical behavior near the end point for 
true events is 
important to establish the end point of the inclusive distribution.  
We find that the inclusive analysis is useful to obtain the information
on the heaviest of the squark/gluino.
 }
\end{center}
\end{titlepage}
\setcounter{footnote}{0}
\section{Introduction}
While the particle interactions at low energy are described correctly by
the standard model (SM),  the mechanism of electroweak  symmetry breaking by
Higgs boson suffers from the fine turning problem. In addition, the
SM
is not successful to describe the dark matter in our Universe.

We expect to obtain information on the physics beyond the standard
model at ATLAS and CMS
experiments at the  LHC, which is scheduled to start this year (2008).
Among  the various proposals, the phenomenology of the
models with quark  and gauge partners with multiplicatively
conserved parity, such as supersymmetric models  with conserved R parity,
Little Higgs models
with T parity and and universal extra dimension models,
get much attention.  In the supersymmetric models,  quark and gluon
partners (squark and gluino)
are pair produced  at the LHC, and subsequently decay into the SM
particles and the lightest supersymmetric particles (LSP's).
The signature at the LHC
will be the high $p_T$ jets and leptons with significant missing
transverse momentum 
which arises from the LSP escaping from detection.
By  the end of the LHC experiment, the squark and gluino in minimal
supergravity model
will be searched up to $\sim$  2.5~TeV
\cite{Atlas,Abdullin:1998pm}.

Interests in  the new physics go beyond the discovery. Many studies
have been carried out to find out the possible clues to study  \lq\lq the
nature of the new physics\rq\rq,
such as the masses, spins and interactions of the new particles.
  The progress has been made especially in  the exclusive
channels.   The end points of the invariant mass distributions
constrain the  sparticle masses and for some cases nearly  all sparticle
masses can be measured. 
The end point study  is extremely successful when the decay involves
many leptons \cite{Atlas,Abdullin:1998pm,Hinchliffe:1996iu,Hinchliffe:1999zc,Bachacou:1999zb,Allanach:2000kt}.

Other important variables are transverse masses.  The peak of the
effective mass
distribution, which
is the sum of the transverse momenta of jets, leptons and
$E_T^\mathrm{miss}$, is related 
to the sum of the sparticle masses produced by $pp$ collision. The
quantity is inclusive and
would be useful in the early stage of the LHC experiment. More
sophisticated quantity is the $m_{T2}$ variable 
\cite{Lester:1999tx,Barr:2003rg}.
This can be calculated from the two visible objects, and the missing momentum
of the events and a test  LSP mass.
Recently, the quantity acquires much of attention because  this
function  has a cusp at
the correct LSP mass when the squark/gluino undergoes  three body
decays.  Some exercises
have been carried out for some  model points where only 
gluino - gluino production can be observed, or for several other decay
patterns without specifying
the selection processes 
\cite{Cho:2007qv,Gripaios:2007is,Barr:2007hy,Cho:2007dh}.

In this paper, we propose an  inclusive study of $m_{T2}$ variables using
a hemisphere
method.  Namely  we group jets into two  ``visible objects''
and calculate $m_{T2}$ variable based on them. 
The grouping algorithm is called a hemisphere method
and discussed in earlier works \cite{hemi,Matsumoto:2006ws}.
 The motivation is to 
collect the cascade decay  products from a squark or a 
gluino enough probability to see the $m_{T2}$ end point,
and obtain
the information on their masses  without going into  exclusive analysis.  
If it works,  the LSP mass also can be obtained  in the early stage  of the
 experiment. 

 We found the correspondence between the reconstructed $m_{T2}$ end
point  and the mass of squark or gluino is good. 
We recognized that $m_{T2}$ is sensitive to 
max($m_{\tilde g} , m_{\tilde q}$)
as $m_{T2}$ is defined from the maximum of transverse mass of the
visible objects for the test LSP momenta 
consistent with   $E_T^\mathrm{miss}$.
Not surprisingly,
the  probability to reconstruct the correct object is rather low and 
the end point is smeared, however,
we find the event-wise response of the $m_{T2}$  to the test LSP mass 
mentioned  in Refs.\cite{Cho:2007qv,Gripaios:2007is,Barr:2007hy,Cho:2007dh}
is  useful  to ensure  the correctness of the end point.

In this paper, we especially compare the mixed modulus anomaly mediation
(MMAM) model to the supergravity (SUGRA) model.
The MMAM predicts a degenerate mass spectrum in some parameter region
of the model, where the sparticles are heavy while 
available 
$p_T$'s of the daughter particles 
are small. This is the model where the kink at the
LSP mass should appear in a rather  high value. 
This point may be compared with 
model points in the mSUGRA (minimal Supergravity) model, 
where the gluino mass is lighter than the 
corresponding MMAM point, but the squark mass is much heavier than
the gluino so that the total squark/gluino production cross sections are
same.
We find that the $m_{T2}$ end point is useful 
to extract the  squark mass for this case, therefore the MMAM and the SUGRA 
models can be distinguished.  

This paper is organized as follows. In Section \ref{sec:mt2} we review
the $m_{T2}$ variable  and  the cusp structure appearing 
in the endpoint of the $m_{T2}$ distribution as the function 
of the test LSP mass. In Section \ref{sec:MC}
we describe the inclusive $m_{T2}$ and 
perform a Monte Carlo simulation to study the
distribution. Section \ref{sec:sum} is devoted 
to the conclusion.

\section{Transverse mass ($m_{T2}$) }
\label{sec:mt2}
In hadron collisions, squarks and gluinos are produced in pair and
these SUSY particles decay subsequently into the final states including
jets, leptons,  and two LSP's. The LSP is usually the lightest neutralino.
With $R$-parity conservation, the LSP is a neutral and stable particle 
and it escapes from detection.
There are two LSP's in the final state and one cannot measure each LSP 
momentum experimentally while the total transverse momentum can
be measured. 
The stransverse mass  $m_{T2}$ is defined as follows:
\begin{eqnarray}
 m^2_{T2}(m_\chi)\equiv 
\min_{\mathbf{p}_{T1}^{\mathrm{miss}}
+\mathbf{p}_{T2}^{\mathrm{miss}}=\mathbf{p}_{T}^{\mathrm{miss}}}
\left[ \mathrm{max}
\left\{
m^2_{T}(\mathbf{p}_{T1}^{\mathrm{vis}},
\mathbf{p}_{T1}^{\mathrm{miss}}),
m^2_{T}(\mathbf{p}_{T2}^{\mathrm{vis}},
\mathbf{p}_{T2}^{\mathrm{miss}})
\right\}
\right],
\label{eq:mt2}
\end{eqnarray}
where $\mathbf{p}^{\mathrm{vis}}_{Ti}$ is the transverse 
momentum of a ``visible object'' from a squark/gluino decay,
which is defined as the sum of visible particle momenta.
The $\mathbf{p}_{T}^{\mathrm{miss}}$ is the total missing transverse
momentum. 
The minimization is taken with respected to the unknown
LSP momenta $\mathbf{p}_{T1}^{\mathrm{miss}}$, 
$\mathbf{p}_{T2}^{\mathrm{miss}}$ under the constraint
${\mathbf{p}_{T1}^{\mathrm{miss}}
+\mathbf{p}_{T2}^{\mathrm{miss}}=\mathbf{p}_{T}^{\mathrm{miss}}}$.
The transverse mass, $m^2_{T}$, is defined as 
\begin{eqnarray}
 m^2_{T}\left(\mathbf{p}_{Ti}^\mathrm{vis},\mathbf{p}_{Ti}^{\mathrm{miss}}
\right)
 =(m_i^{\mathrm vis})^2+m_\chi^2+2\left(E_{Ti}^{\mathrm {vis}}E_{Ti}^{\mathrm miss}
-\mathbf{p}_{Ti}^{\mathrm {vis}}\cdot\mathbf{p}_{Ti}^{\mathrm{miss}}
\right),
\end{eqnarray}
where $E_{Ti}=\sqrt{p_{Ti}^2+m^2_{\chi}}$.
It should be noted that the true LSP mass $(m_{\chi_1^0})$ is unlikely 
to be known in advance, 
so $m_{T2}$ is regarded as a function of a test LSP mass ($m_\chi$). 

One of the important features is that the $m_{T2}$ is smaller
than the parent gluino/squark masses if  a test LSP mass is set equal 
to the true value. 
\begin{eqnarray}
 m_{T2}(m_{\chi^0_1})\leq \mathrm{max}(m_{\tilde q},m_{\tilde g}).
\end{eqnarray}
From the upper end point 
of $m_{T2}$ ($m_{T2}^\mathrm{max}$), one can obtain
the information on the mass of the parent particle.
Without knowledge of the true LSP mass, $m_{T2}^\mathrm{max}$
provides a one-dimensional constraint between the masses of
the squark/gluino and the LSP.

Recently, it is pointed out that the $m_{T2}^\mathrm{max}(m_{\chi})$ 
function
has a kink structure at which $m_{\chi}$ is the true LSP mass unless 
the squark/gluino decays directly into the LSP through a two body decay. 
An analytic expression of $m_{T2}^\mathrm{max}$ is derived
in Refs. \cite{Barr:2007hy,Cho:2007dh}.
If one considers events in which the squark and  the gluino are
produced in pair with a vanishing total transverse momentum,
$m_{T2}^\mathrm{max}(m_{\chi})$ is given as follows.
\begin{eqnarray}
\label{eq:mt2max}
m_{T2}^\mathrm{max}(m_{\chi})=
\left\{\begin{array}{ll} {\cal F}^{\rm
max}_<(m_{\chi})
  ~~~ \mbox{for} ~~
m_{\chi}<m_{\chi_1^0} \\ 
{\cal F}^{\rm max}_{>}(m_{\chi})
~~~
\mbox{for} ~~ m_{\chi}>m_{{\chi}_1^0},
\end{array}\right.
\end{eqnarray}
where
\begin{eqnarray}
\label{eq:func}
 {\cal F}^{\rm max}_< (m_{\chi})&=&
{\cal F}(m^{\mathrm{vis}}_{1}=m^{\mathrm{vis}}_{\mathrm{ min}}, 
m^{\mathrm{vis}}_{2}=m^{\mathrm{vis}}_{\mathrm{min}},
\theta=0,m_\chi),\nonumber 
\\
{\cal F}^{\mathrm{max}}_{>} (m_{\chi})&=&{\cal F}
(m^{\mathrm{vis}}_{1}=m^{\mathrm{vis}}_{\mathrm{max}},
 m^{\mathrm{vis}}_{2}=m^{\mathrm{vis}}_{\mathrm{max}},\theta=0,m_\chi).
\end{eqnarray}
Here the function ${\cal F}$ is given in Ref.\cite{Cho:2007dh} and
$m^{\mathrm{vis}}_{i}$ is kinematically bounded as follows,
\begin{eqnarray} 
m^{\mathrm{vis}}_{\mathrm {min}}\leq m^{\mathrm{vis}}_{i}\leq
m^{\mathrm{vis}}_{\mathrm{ max}}.
\end{eqnarray}
Notice that the events at the end point satisfy  
$m^{\mathrm{vis}}_{i}=m^{\mathrm{vis}}_{\mathrm {min}}$
for $m_{\chi}<m_{\chi_{1}^{0}}$ while 
$m^{\mathrm{vis}}_{i}=m^{\mathrm{vis}}_{\mathrm {max}}$
for $m_{\chi}>m_{\chi_{1}^{0}}$. 
The kink structure of $m_{T2}^\mathrm{max}(m_\chi)$ appears 
 since the functional form of $m_{T2}^\mathrm{max}(m_\chi)$ changes at 
$m_{\chi}=m_{\chi_{1}^0}$.
In Ref.\cite{Barr:2007hy}, it is shown that the kink structure 
appears even if the pair-produced squark and gluino have a non-vanishing
transverse momentum. 
If one can identify the position of the kink from the LHC experiment, 
one can determine the masses of the squark/gluino and the LSP simultaneously. 
In Ref. \cite{Cho:2007dh}, 
it is demonstrated that the masses of the squark/gluino and
the LSP are determined using  exclusive decay channel by performing
Monte Carlo simulations.
In particular, for the case that
the gluino decay 
${\tilde g}\to qq \chi_1^0$ occurs through the off-shell squark
exchange diagram, the $m_{T2}^\mathrm{max}$ from the gluino pair 
production has a very  sharp kink structure
and the masses are determined precisely.

\section{Inclusive $m_{T2}$  Analysis}
\label{sec:MC}
\subsection{Hemisphere analysis and inclusive $m_{T2}$ parameter}
\label{sec:hemi}
In this section we argue that the kink method discussed in
Sec.\ref{sec:mt2} can be extended to an inclusive analysis.
 In case of exclusive analyses, one needs to specify
a cascade decay chain.  The branching ratio of the
cascade decay chain would depend on the model parameters.
On the other hand,  inclusive distributions are 
rather insensitive to branching ratios.
Therefore, if an inclusive quantity can be defined,
it may be useful to determine the squark and the gluino masses
in the early stage of the LHC experiment.
One disadvantage of  inclusive approaches  may be that all the production
and decay modes should be  taken into account simultaneously,
 and  $m_{T2}$ distribution may be smeared.

To define an inclusive  $m_{T2}$ distribution, we 
group the final particles into two ``visible objects''. 
For this purpose, we adopt a hemisphere method in Ref. \cite{hemi,
Matsumoto:2006ws}.
For each event,  two hemispheres are defined and
high $p_{T}$ jets, leptons, and photons are assigned into
one of the hemispheres as follows;
\begin{enumerate}
 \item Each hemisphere is defined by an axis
       $p^\mathrm{vis}_{i}(i=1,2)$,  which is the
       sum of the momenta of high $p_{T}$ objects belonging to
       hemisphere $i$. We require $p_{T}>50$ GeV for jets 
       to reduce QCD backgrounds.
\item High $p_{T}$ objects $k$ belonging to hemisphere $i$
      satisfy the following conditions:
      \begin{eqnarray}
       d(p_i,p^\mathrm{vis}_{i}) < d(p_k,p^\mathrm{vis}_{j}),
      \end{eqnarray}
      where the function $d$ is defined by
      \begin{eqnarray}
       d(p_k,p^\mathrm{vis}_{i}) 
      =(E_i-|p^\mathrm{vis}_{i}|\cos\theta_{ik})\frac{E_i}{(E_i+E_k)^2}.
      \end{eqnarray}
      Here $\theta_{ik}$ is the angle between $P_i$ and $p_k$.
\end{enumerate}
To find axises $p_i^\mathrm{vis}$, we adopted the algorithm discussed
in Ref.\cite{hemi, Matsumoto:2006ws}.
Once $p_i^\mathrm{vis}$'s are determined,  one can calculate 
$m_{T2}$  by using Eq.(\ref{eq:mt2}).

The inclusive $m_{T2}$ may be compared with MTGEN \cite{Lester:2007fq}.
The MTGEN variable is a minimum of the $m_{T2}$ variable 
for all possible choices of 
two subsets of particles $\alpha$ and $\beta$.  The correct choice of subsets
$\alpha$ and $\beta$ leads  the heaviest sparticle mass as the end point,
and the end point should be bounded from above by the mass, if the
initial state radiation can be ignored.  In our algorithm, we assume that
the algorithm described above groups high $p_T$ jets from the same cascade
decay with enough probability. This approach
is useful if the visible hemisphere mass is small compared with the
the leading jet energies, which is expected especially 
for the events near the $m_{T2}$ end 
point when the test mass is smaller than the LSP mass, 
see Eqs.(\ref{eq:mt2max}), (\ref{eq:func}).

\subsection{Model points}
\label{sec:model}
To perform a Monte Carlo analysis,  we choose two sample points, A and B.
The point A corresponds to the MMAM  model 
\cite{Choi:2004sx,Choi:2005ge,Choi:2005uz,Endo:2005uy}.
In the MMAM model, the mass spectrum is parametrized by the 
modular weights for matter fields $n_i$ and the gravitino mass ($m_{3/2}$) 
and $R\equiv m_{3/2}\langle (T+T^*)/F_T \rangle$ where $T$ and $F_T$
are a modulus field  and its $F$-component, respectively. 
In general, the MMAM model predicts a degenerate SUSY spectrum
compared with the mSUGRA  model. 
If $\alpha =R/\ln(M_\mathrm{pl}/{m_{3/2}})$
 is large, the SUSY spectrum becomes more degenerate.
In this analysis, we choose the point studied in Ref.\cite{Kawagoe:2006sm},
$n_i=0(1)$ for squarks and sleptons (Higgs boson),
$R=20$, $\tan\beta=10$ and 
the gravitino mass is determined so  that $M_3=650$~GeV at the GUT scale.
The point B corresponds to the mSUGRA with $m_0=1475$~GeV, 
$m_{1/2}=561$~GeV, $A=0$ and  $\tan\beta=10$.

\begin{table}
\begin{center}
\begin{tabular}{|c|c|c|}
\hline
   & A: MMAM  &  B: mSUGRA    \\
\hline
   & $n_i=0$, $R=20$, 
   & $m_0=1475$, $m_{1/2}=561.2$, \\
   & $M_3(\mathrm{GUT})=650$ 
   & $A=0$, $\tan\beta=10$      \\
\hline
\hline
${\tilde g}$ & 1491&  1359   \\
${\tilde u}_L$ & 1473& 1852   \\
${\tilde u}_R$ & 1431& 1831   \\
${\tilde d}_R$ & 1415& 1830   \\
${\tilde \chi}^0_1$ & 487 & 237   \\
\hline
\end{tabular}
\caption{Relevant SUSY mass parameters at points A and B.
All the mass parameters are given  in GeV.
\label{tab:mass} }
\end{center}
\end{table}

The mass spectrum of SUSY particles is calculated using
ISAJET \cite{Paige:2003mg} for each sample point. 
In Table.{\ref{tab:mass}}, the relevant SUSY 
masses are listed. At point B, $m_{\tilde q}>m_{\tilde g}$ 
and the gluino undergoes three-body decay through the
off-shell squark diagram. 
The total production cross section  of SUSY events at the LHC is
$\sigma = 0.13$ pb for both points.  
Squark-gluino coproduction  is larger than squark-squark and gluino-gluino 
productions for both points.

The point B is chosen so that the $M_\mathrm{eff}$ 
distribution of one lepton mode is very similar to that for point A, 
where $M_\mathrm{eff}$  is defined from the sum of the $p_T$ of the 
first four jets and a lepton and the missing transverse momentum as follows,
\begin{eqnarray}
\label{eq:meff}
 M_{\mathrm{eff}}=\sum_{i=1}^4 p_{Ti} +p_{Tl}+ E_{T}^\mathrm{miss}.
\end{eqnarray}
For the Monte Carlo analysis, we generate $5\times 10^4$ SUSY events 
by HERWIG 6.5  \cite{Corcella:2000bw}
for each sample point.  To estimate event
distributions measured by the LHC detector, we use AcerDET
\cite{RichterWas:2002ch}. This code provides a simple detector
simulation at the LHC.

\begin{figure}[htb]
    \centerline{
 \epsfxsize=\figscale\textwidth\epsfbox{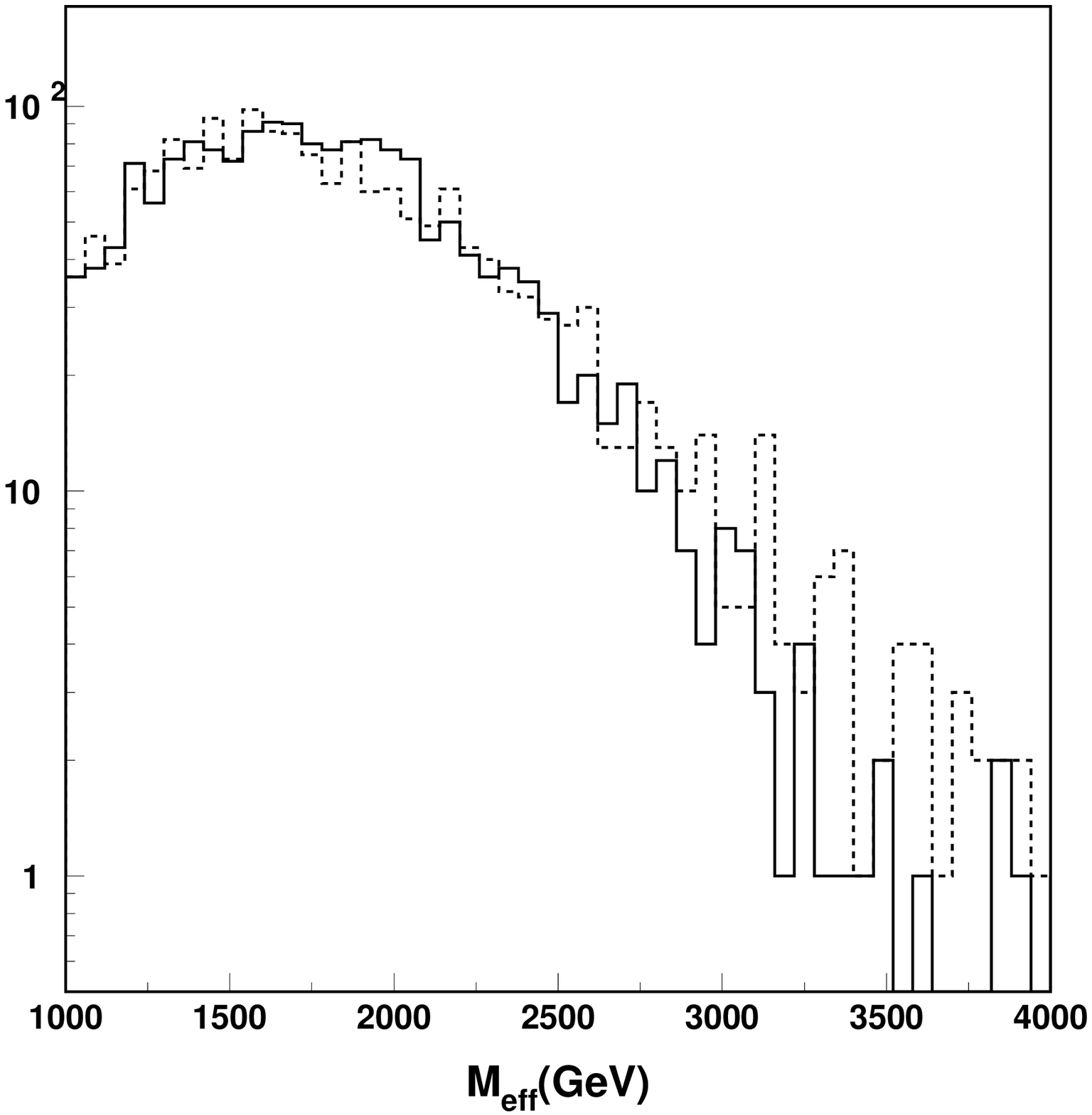}
 \epsfxsize=\figscale\textwidth\epsfbox{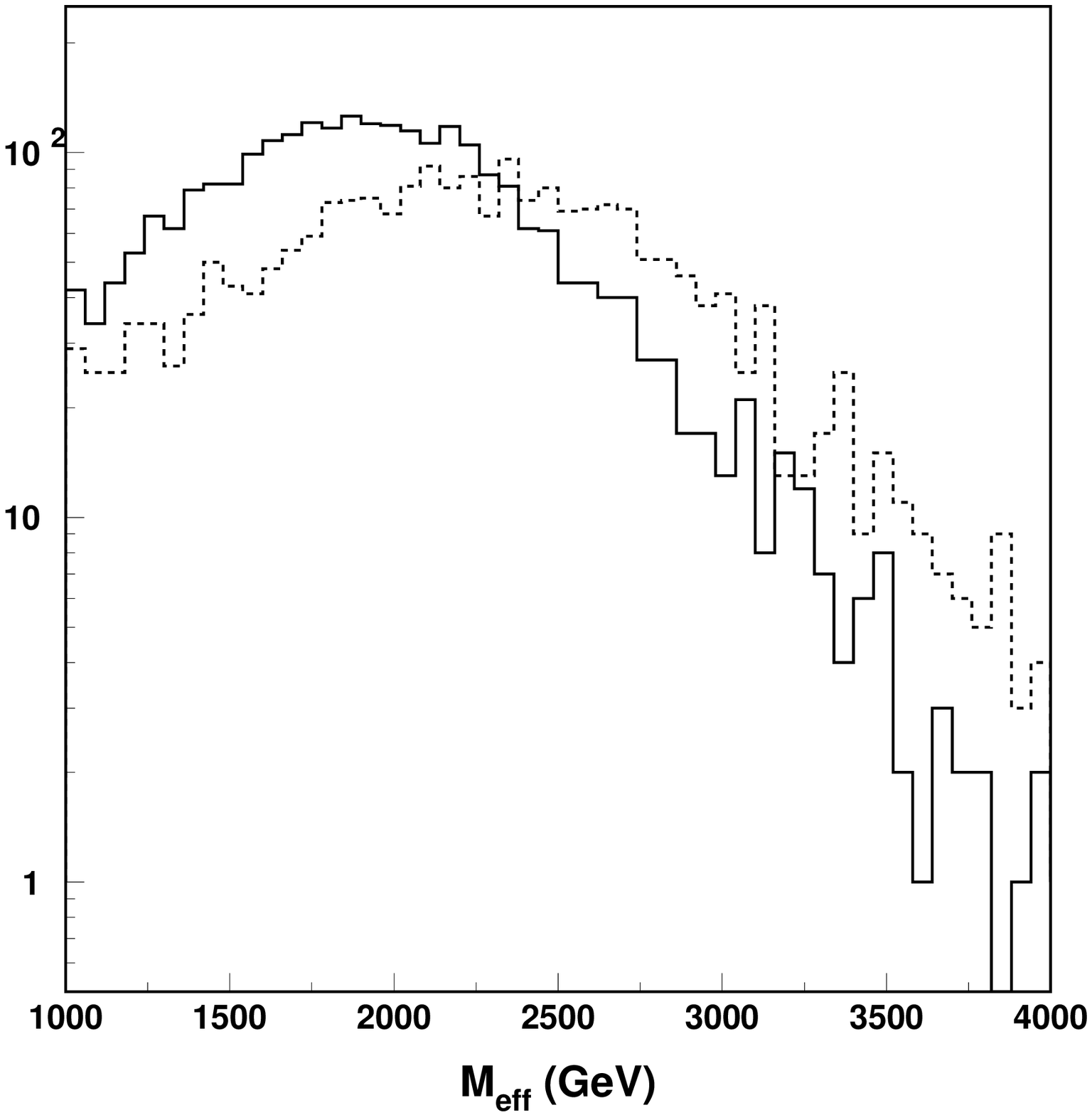}
}
    \caption{(a). $M_{\mathrm{eff}}$ distributions for one lepton channel.
The solid (dashed) histogram is for point A (B).
(b). $M_{\mathrm{eff}}$ distributions for one lepton channel as in (a)
but summing up all jets with $p_T>$~50 GeV.
The solid (dashed) histogram is for point A (B).}
    \label{fig:meff}
\end{figure}

In Fig.\ref{fig:meff}(a), the $M_{\mathrm{eff}}$ distribution is shown
for one lepton channel. Here we require the the following cut.
\begin{enumerate}
 \item $n_\mathrm{jet}(p_T>100~\mathrm{GeV})\equiv n_{100}\ge$ 1
       and $n_\mathrm{jet}(p_T>50~\mathrm{GeV})\equiv n_{50}\ge$ 4
       within $|\eta|<3$. 

\item $E_T^\mathrm{miss}>0.2 M_\mathrm{eff}$ and $E_T^\mathrm{miss}>
      100$ GeV and $S_T>0.2$.

\item There is one isolated lepton with $p_T>20$~GeV.

\end{enumerate}
The solid (dashed) histogram is the distribution
for point A (B) and 
the $M_{\mathrm{eff}}$ distributions roughly agree.

Although there is not much difference in 
 $M_{\rm eff}$ distribution defined in Eq.(\ref{eq:meff}),
there are  more high $p_T$ jets  on average at point B compared with point A. 
This is because the squark-gluino coproduction is dominant, 
and a squark decaying into a gluino  leads additional high
pt jets in the events.
If one sums all jets with $p_T>$50~GeV, then the distribution
of point B is significantly higher than that of point A.
In Fig.\ref{fig:meff}(b), the $M_{\mathrm{eff}}$ distribution summing
up all jets with $p_T>$ 50~GeV is shown for one lepton channel. 
We will see in the  next subsections that inclusive $m_{T2}$ analyses 
give us a more quantitative measure to the difference of the two points. 

\subsection{Monte Carlo analysis: Point A (MMAM)}

First, let us consider point A. 
We require the following cut to select the events.
\begin{enumerate}
 \item $n_{100}\ge 2$ 
       and $n_{50}\ge 4$ 
       within $|\eta|<3$. 

\item  The effective mass of the event must satisfy
      $M_{\mathrm{eff}}>1200$~GeV.

\item At least two jets in each hemisphere. 

\item $E_T^\mathrm{miss}>0.2 M_\mathrm{eff}$ and $E_T^\mathrm{miss}>
      100$~GeV.

\item There is no isolated lepton with $p_T>20$~GeV.

\end{enumerate}
With these cuts, the standard model backgrounds are expected to be 
reduced significantly, so we  do not consider the SM background 
in this simulation.

\begin{figure}[htb]

    \centerline{
 \epsfxsize=\figscale\textwidth\epsfbox{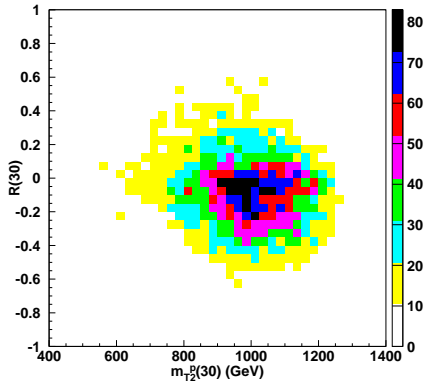}
}

    \caption{The two-dimensional distribution 
in the  $m^{(p)}_{T2}$-$R$ plane for point A. 
The test LSP mass is assumed as $m_{\chi}=30$ GeV.}
    \label{fig:ratio_kklt}
\end{figure}

To check how well the hemisphere method works, let us consider the
following ratio.
\begin{eqnarray}
R(m_\chi)\equiv 
\frac{m_{T2}(m_\chi)-m_{T2}^{(p)}(m_\chi)}{m_{T2}^{(p)}(m_\chi)}.
\end{eqnarray}
where the parton level $m_{T2}^{(p)}$ is 
defined  so that  each visible momentum
is the difference of a initially produced sparticle and the daughter 
LSP momentum, and $E_{T}^{\mathrm {miss}}$ is the missing energy 
provided by AcerDET after smearing.
 If the hemisphere method works well,
$R\sim 0$. In Fig.\ref{fig:ratio_kklt},
the two-dimensional distribution 
in the $m^{(p)}_{T2}$-$R$ plane is shown for $m_\chi=30$~GeV. 
The peak of the distribution appears around  $R\sim 0$,
but the deviation can be large.
The reconstructed $m_{T2}$ tends to be  smaller than $m_{T2}^{(p)}$.
The main source of the deviation is the mis-grouping
of the visible objects under the hemisphere method, and 
also neutrinos and jets with $p_T<50$~GeV which are not included 
in the hemisphere.  

\begin{figure}[htb]
    \centerline{
\epsfxsize=\figscale\textwidth\epsfbox{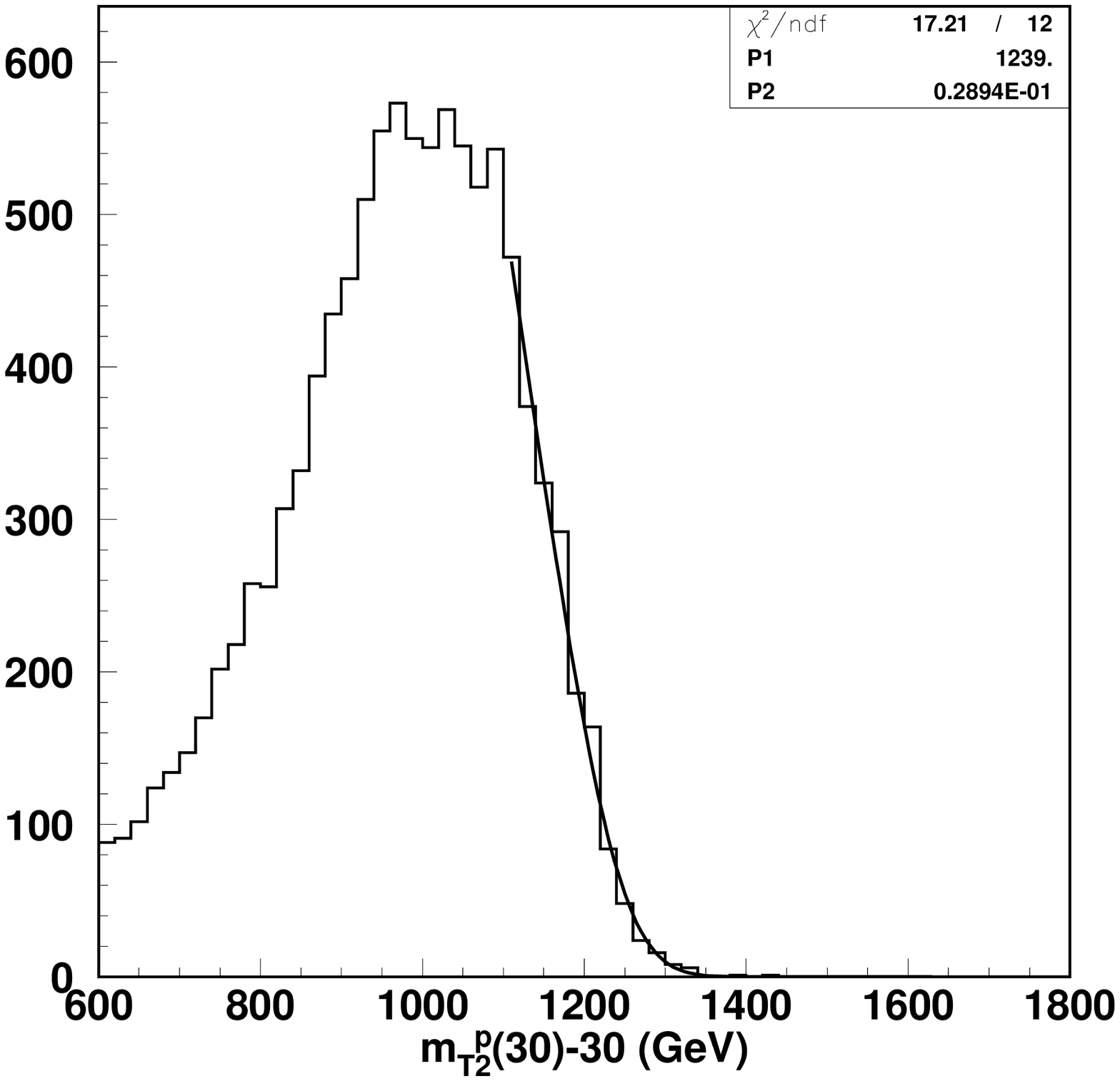}
\epsfxsize=\figscale\textwidth\epsfbox{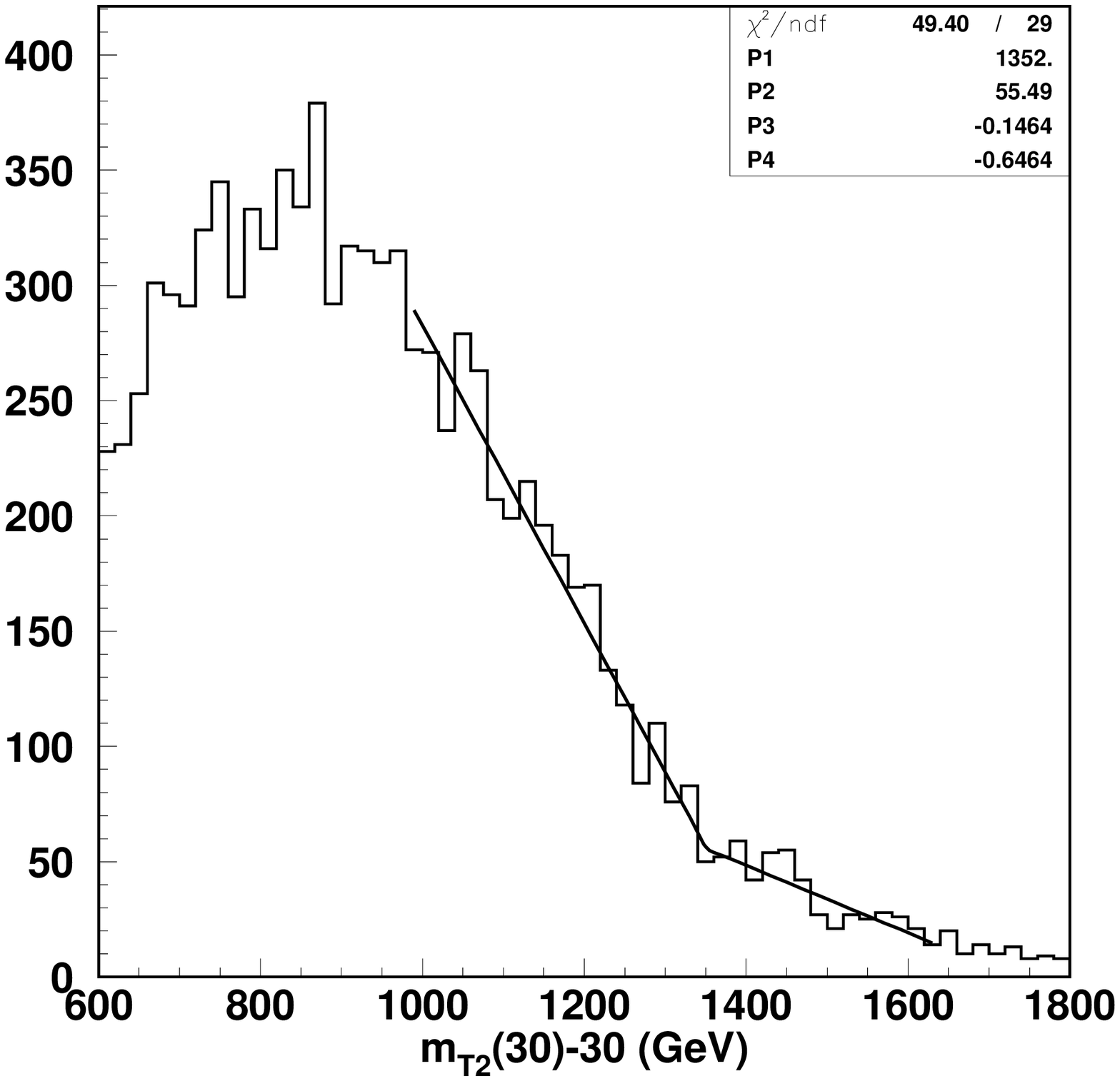}
}
    \caption{(a). The $m_{T2}-m_{\chi}$ distribution at parton level for
 $m_\chi=30$~GeV.
(b). The reconstructed $m_{T2}-m_{\chi}$ distribution for  $m_\chi=30$~GeV.
 Fitting functions of the end points are also shown, see text.   }
    \label{fig:mt2_kklt_30}
\end{figure}

Let us consider the $m_{T2}$ distribution for $m_{\chi}<m_{\chi_1^0}$.
In Fig.\ref{fig:mt2_kklt_30}(a), the $m_{T2}^{(p)}$ distribution 
is shown for $m_\chi=30$~GeV.
There is an end point at $m_{T2}^{(p)}\simeq 1250$~GeV.
In Fig.\ref{fig:mt2_kklt_30}(b), the reconstructed $m_{T2}$ distribution
is shown for $m_\chi= 30$~GeV.
Compared with the $m_{T2}^{(p)}$ distribution, there is a long tail
due to the mis-grouping of the hemisphere method, although 
there is some structure at $m_{T2}\sim 1250$~GeV. 

\begin{figure}[htb]
    \centerline{
\epsfxsize=\figscale\textwidth\epsfbox{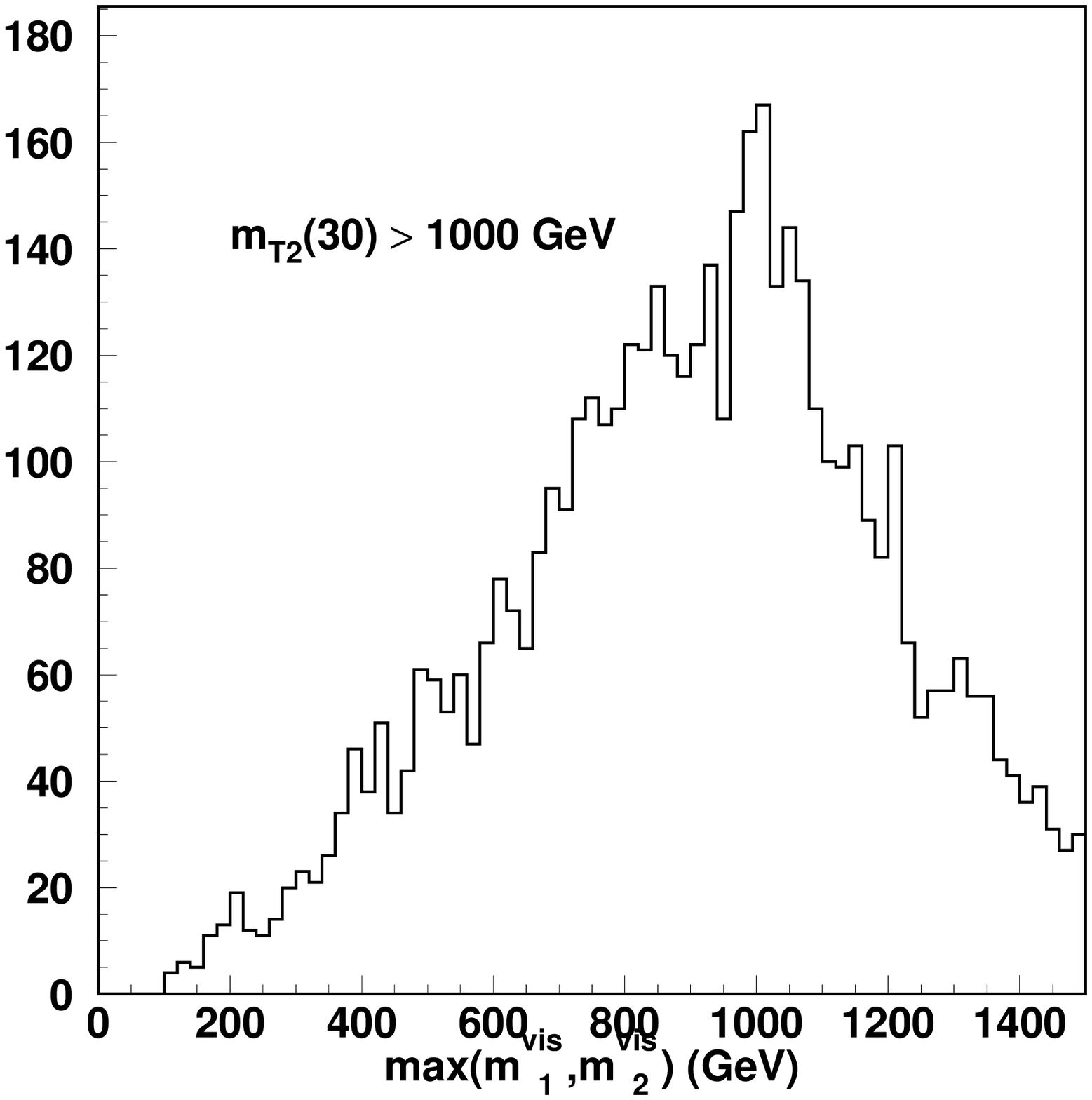}
\epsfxsize=\figscale\textwidth\epsfbox{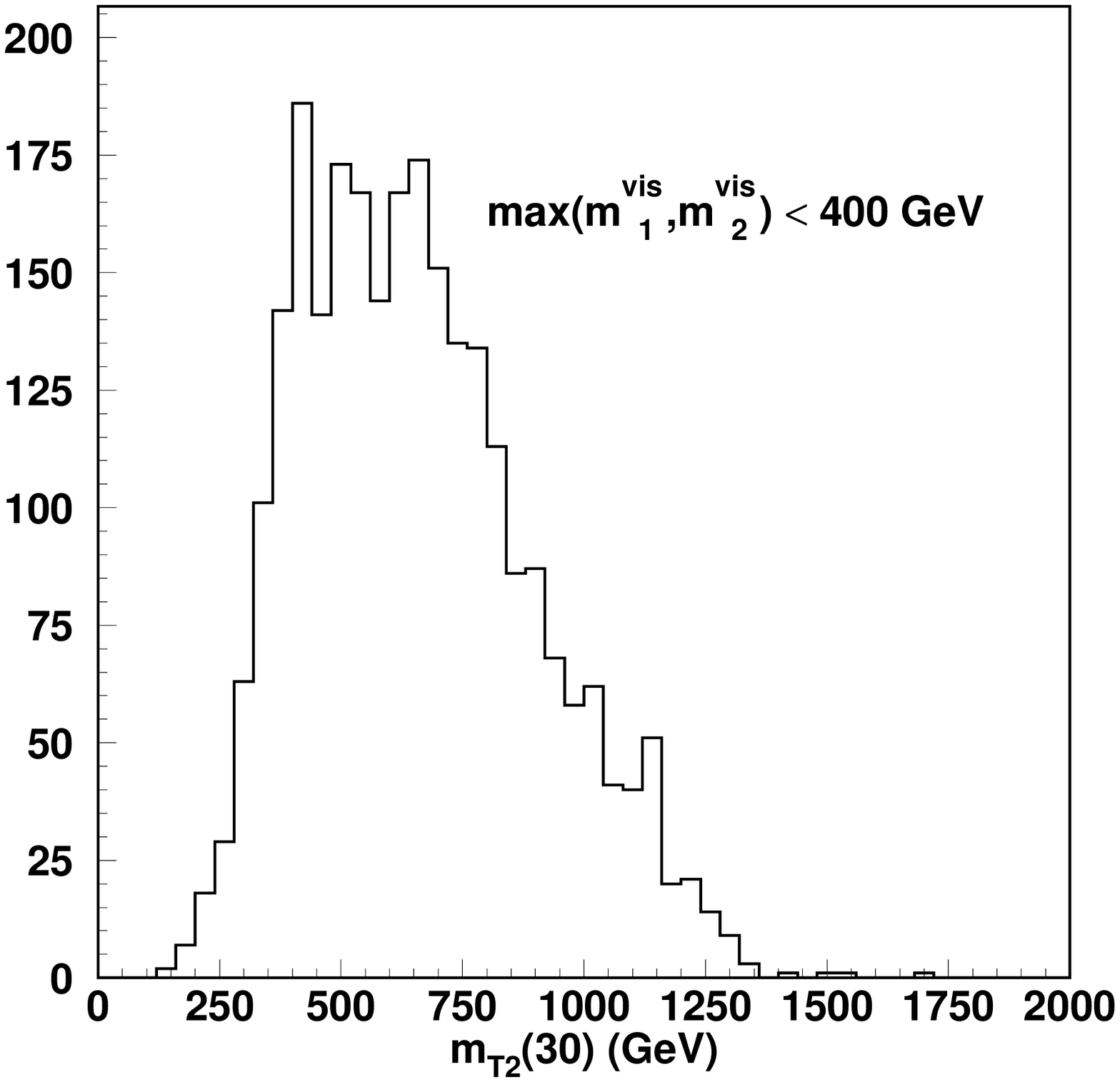}
}
    \caption{(a). The distribution of
 max$(m^{\mathrm{vis}}_1,m^{\mathrm{vis}}_2)$ for 
$m_{T2}(30)>1000$~GeV.
(b). The $m_{T2}(30)$ distribution for
max$(m^{\mathrm{vis}}_1,m^{\mathrm{vis}}_2)<400$~GeV. 
   }
    \label{fig:hem_kklt}
\end{figure}

Let us examine the events around the end point region in detail.
Fig.\ref{fig:hem_kklt}.(a) shows max$(m_1^{\mathrm{vis}},m_2^{\mathrm{vis}})$
distribution for $m_{T2}(30)>1000$~GeV. As discussed in Section \ref{sec:mt2},
the true end point event is realized for 
$m_i^{\mathrm{vis}}=m_{\mathrm{min}}^{\mathrm{vis}}$
and the events with large max$(m_1^{\mathrm{vis}},m_2^{\mathrm{vis}})$ are
considered as  fake events. 
To reduce them, the  $m_{T2}$ distribution for 
max$(m_1^{\mathrm{vis}},m_2^{\mathrm{vis}})<400$~GeV
is plotted in Fig.\ref{fig:hem_kklt}.(b).
With the cut on the hemisphere mass, the long tail of $m_{T2}$ disappears  
and one can see a rather clear end point at $m_{T2}\sim  1350$~GeV.

\begin{figure}[htb]
    \centerline{
\epsfxsize=\figscale\textwidth\epsfbox{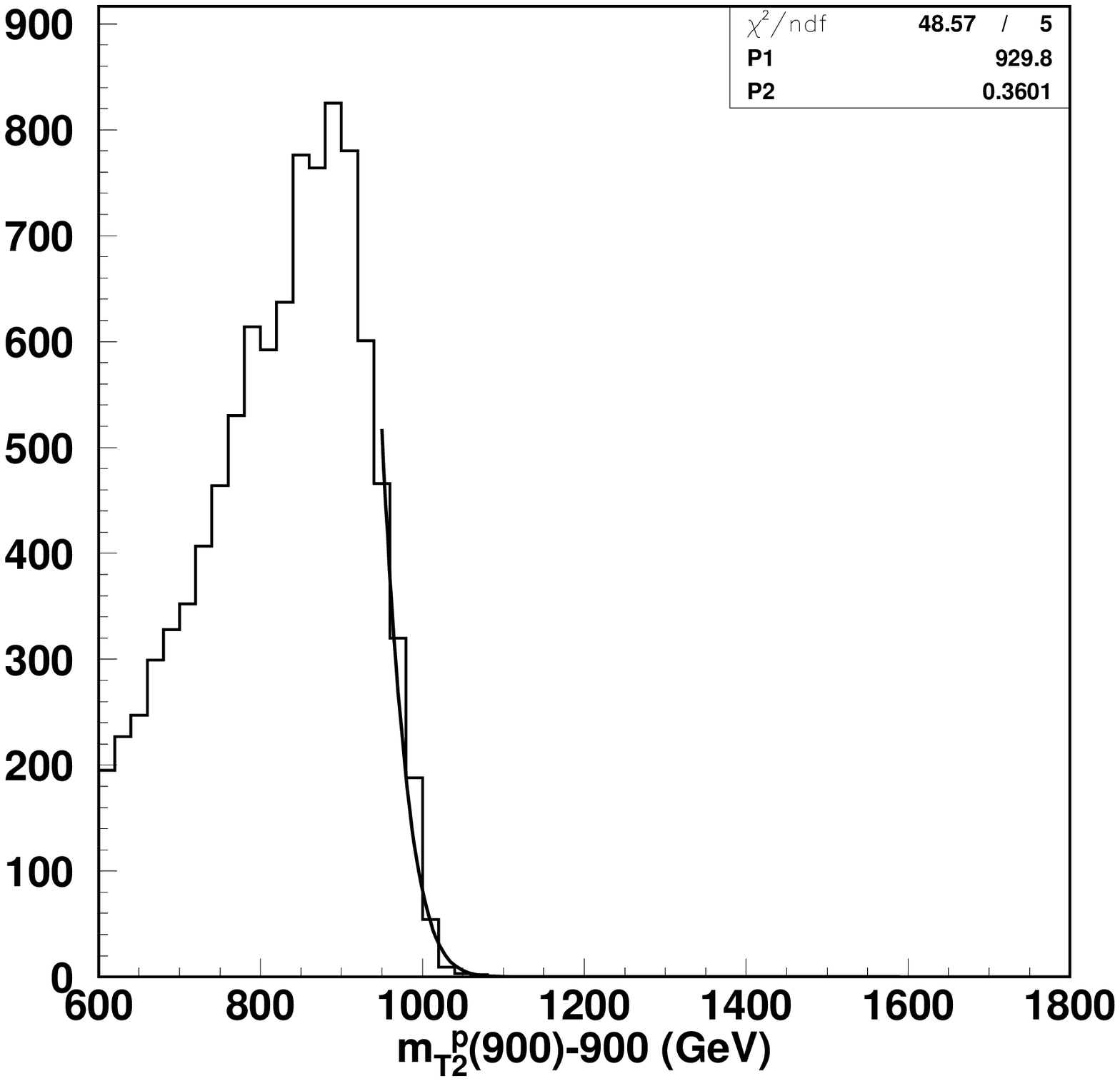}
\epsfxsize=\figscale\textwidth\epsfbox{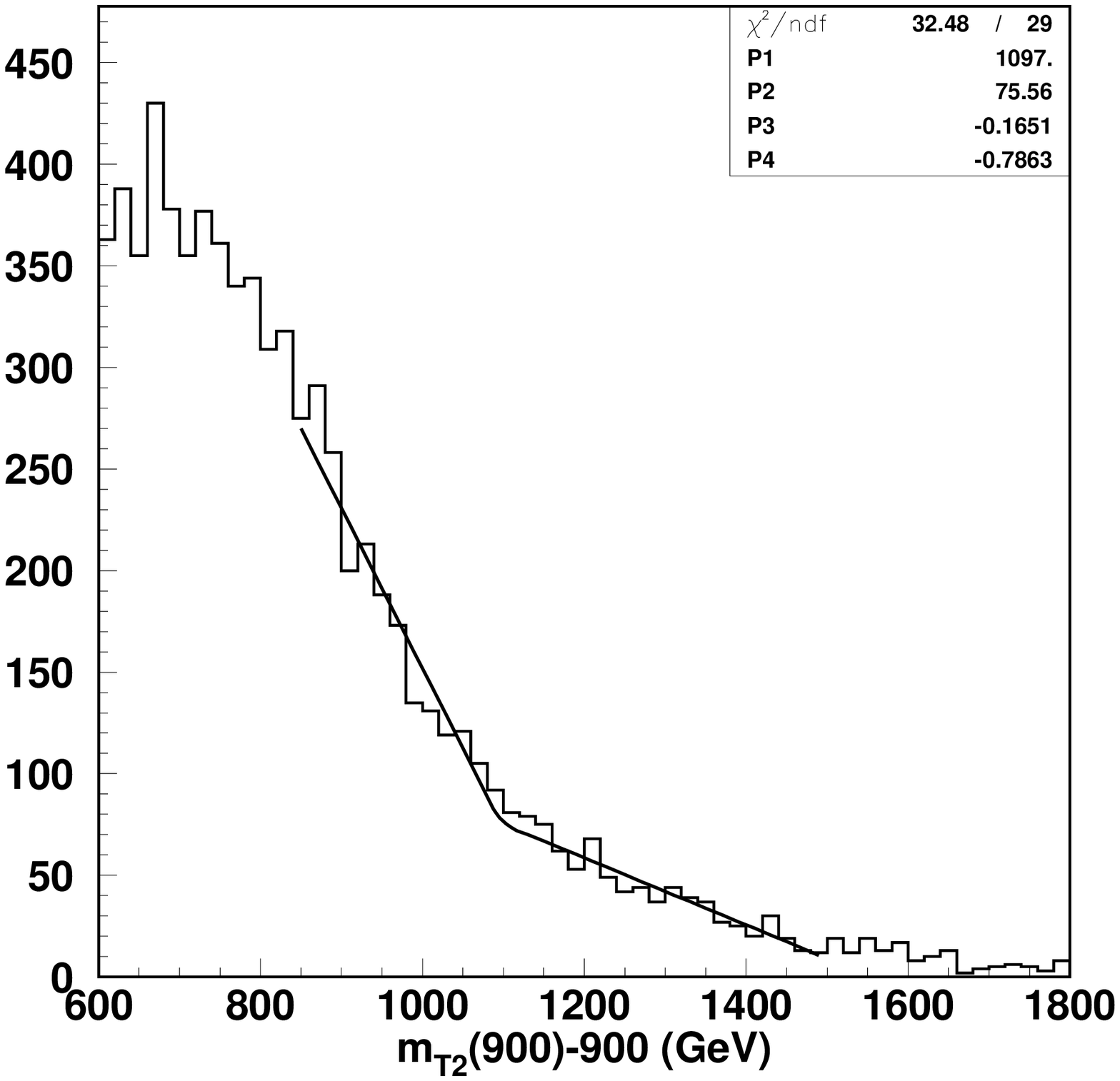}
}
    \caption{(a). The $m_{T2}-m_{\chi}$ distribution at parton level for
$m_\chi=900$~GeV.
(b). The reconstructed $m_{T2}-m_{\chi}$ distribution
for $m_{\chi}=900$~GeV. Fitting functions of the end points 
are also shown, see text. 
   }
    \label{fig:mt2_kklt_900}
\end{figure}

Next, let us consider the $m_{T2}$ distribution for $m_{\chi}>m_{\chi_1^0}$.
In Fig.\ref{fig:mt2_kklt_900} (a), the $m^{(p)}_{T2}-m_{\chi}$ distribution is
plotted for $m_\chi=900$~GeV. There is an end point  at
$m^{(p)}_{T2}\sim1900$~GeV. 
In Fig.\ref{fig:mt2_kklt_900} (b), the reconstructed $m_{T2}-m_{\chi}$ distribution is
plotted for $m_\chi=900$~GeV. The distribution has a long tail 
and one cannot see a clear end point.

\begin{figure}[htb]
    \centerline{
\epsfxsize=\figscale\textwidth\epsfbox{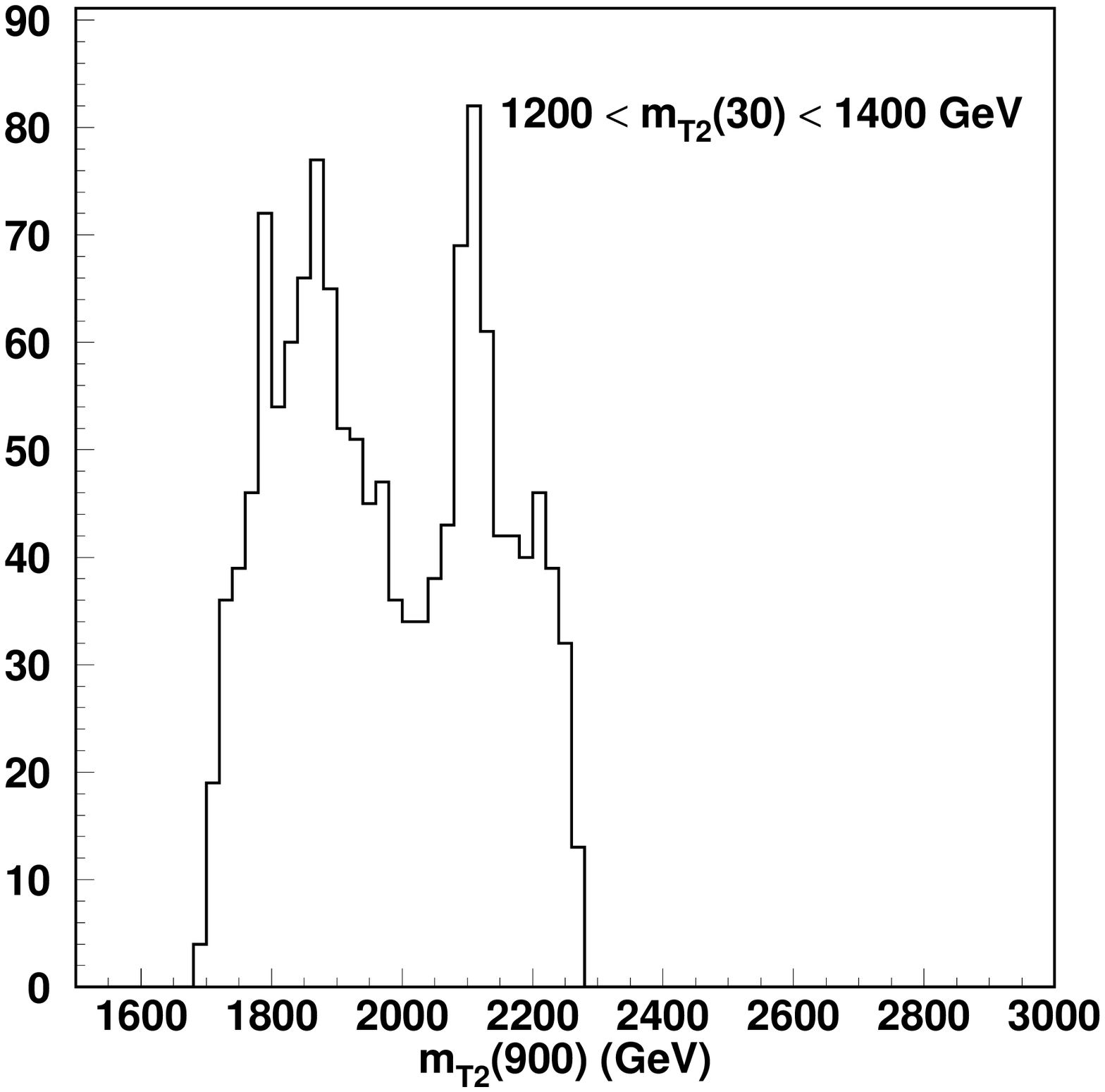}
\epsfxsize=\figscale\textwidth\epsfbox{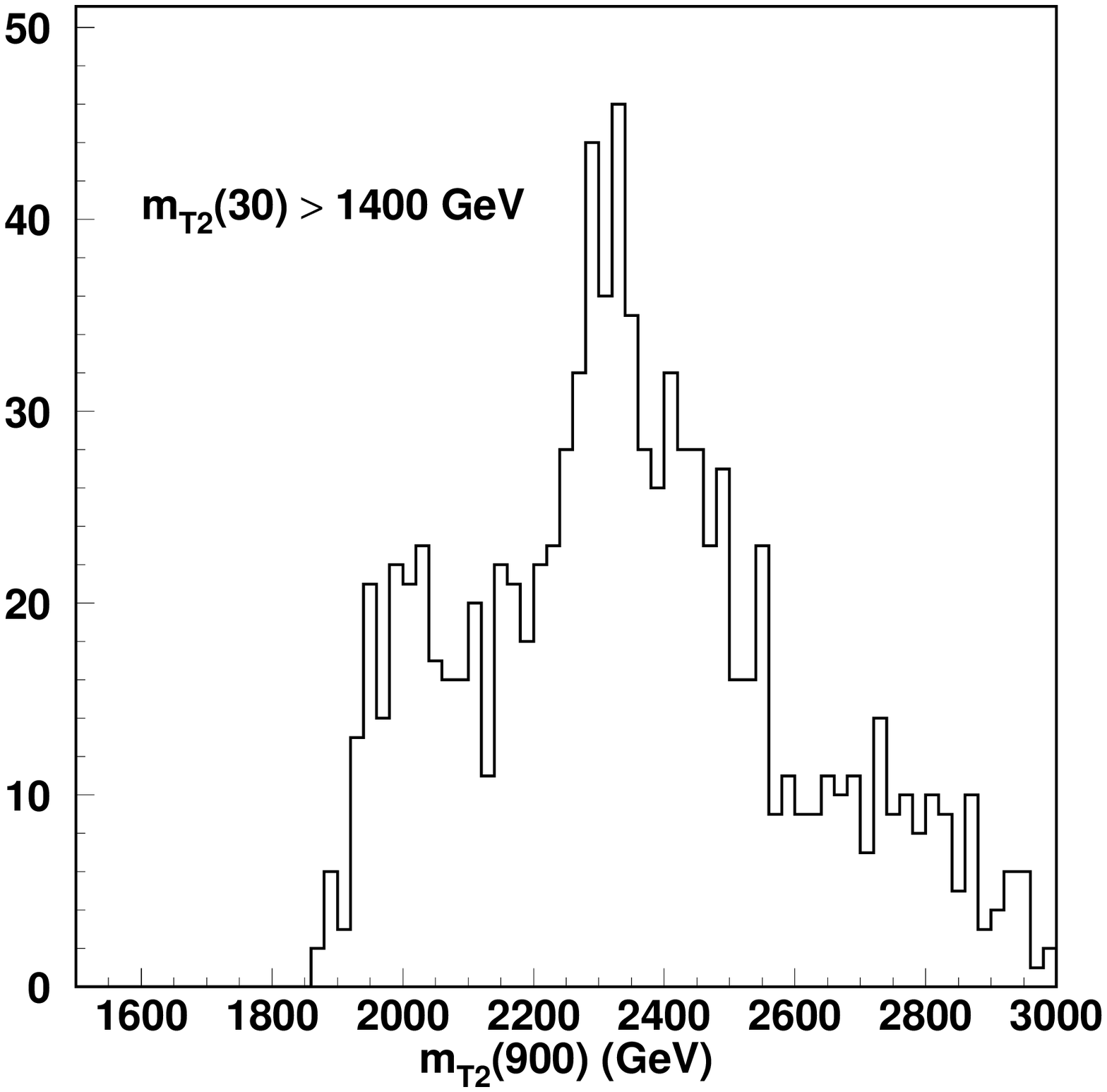}
}
    \caption{(a). The  $m_{T2}(900)$ distribution  
for $1200$~GeV$< m_{T2}(30)< 1400$~GeV (left).
(b). The $m_{T2}(900)$ distribution  
for the fake events $m_{T2}(30)> 1400$~GeV (right).
   }
    \label{fig:mt2_kklt_900a}
\end{figure}

As discussed in Sec.\ref{sec:mt2}, the cusp structure of
$m_{T2}^{\mathrm max}$ appears    since 
the functional form $m_{T2}^{\mathrm{max}}(m_\chi)$ changes at 
$m_\chi=m_{\chi_1^0}$.
The end point event for $m_\chi<m_{\chi_1^0}$ is different from the 
one for $m_\chi>m_{\chi_1^0}$ and these end point events are
 interchanged at  $m_\chi=m_{\chi_1^0}$. 
To confirm this, let us consider how the  events near the end point  for 
$m_{T2}(30)$ behaves when $m_\chi$ is large.
In Fig.\ref{fig:mt2_kklt_900a}(a),  the  $m_{T2}(900)$
distribution is plotted for 1200~GeV$<m_{T2}(30)<$1400~GeV.
There are two peaks in the distribution. 
The lower peak is smaller than the true end point $m_{T2}(900)\simeq$
1900~GeV. These events are true end point events 
of $m_{T2}(30)$ while  
the events around the higher peak are fake events.
In Fig.\ref{fig:mt2_kklt_900a}(b),  the  $m_{T2}(900)$
distribution is plotted using events above the true end point
 $m_{T2}(30)>1400$~GeV.  We find no peak lower 
 than 1900~GeV as expected,  because they are fake events 
for $m_{T2}(30)$. 

\begin{figure}[t]
    \centerline{{\epsfxsize=0.5\textwidth\epsfbox{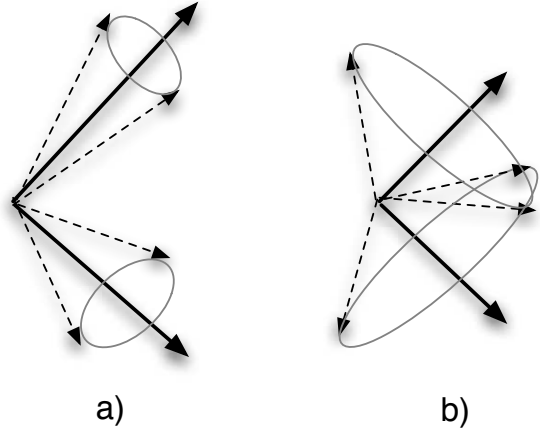}
}}
\caption{
{ Kinematical configurations for 
(a). $m^{\rm vis}\sim m^{\rm vis}_{\rm min}$ and
(b). $m^{\rm vis}\sim m^{\rm vis}_{\rm max}$. 
When $m^{\mathrm{vis}}$ is large, jets in the hemisphere are less
 collinear, and the hemisphere analysis likely misgroups the particles.
}}
\label{fig:hem}
\end{figure}

To find the end points of the $m_{T2}$ distributions,
we also show the fitting of the distribution in 
Figs.\ref{fig:mt2_kklt_30} and \ref{fig:mt2_kklt_900}.
We fit the reconstructed  $m_{T2}$ distribution
with a linear function which changes the slope at some $m_\chi$.
For comparison, we also show the fitting of $m_{T2}^{(p)}$. 
We use a Gaussian smeared fitting function
in Ref. \cite{Bachacou:1999zb} for that.  $\chi^2$/n.d.f is not 
good for both fits,  therefore our fits should be regarded as 
 crude estimates.  In addition, the end point  
 for $m_{\chi}=900$~GeV  depends on the bins used 
 for the fit.  Note that  the end point for $m_{\chi}>m_{\tilde\chi^0_1}$ 
 is realized for the events with $m^{\rm vis}\sim m^{\rm vis}_{\rm max}$, while 
 the efficiency to assign the particles correctly in hemisphere should be 
 low in such case, see Fig.\ref{fig:hem}.

\begin{figure}[htb]
    \centerline{{
\epsfxsize=0.45\textwidth\epsfbox{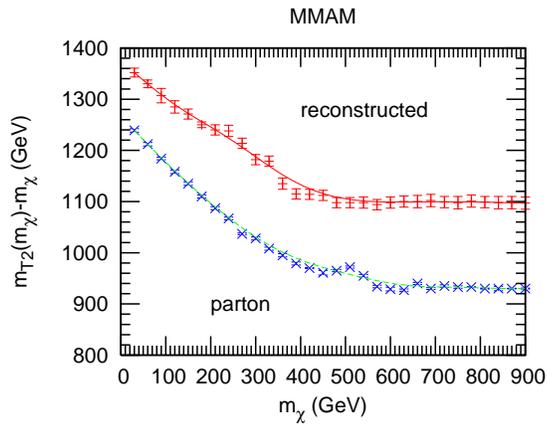}
}}
    \caption{The end point of $m_{T2}(m_\chi)-m_\chi$ 
 for various test LSP masses. The  solid line is
 the  $m_{T2}^\mathrm{max}$  while the dashed line
 is the parton level $m_{T2}^{(p)\mathrm{max}}$. 
   }
    \label{fig:mt2max_kklt}
\end{figure}

In Fig.\ref{fig:mt2max_kklt}, the end points of $m_{T2}$ for 
various test LSP masses are plotted with solid lines. 
The end points of the  $m_{T2}$ are larger
than $m_{T2}^{(p)}$ by $150-200$~GeV, 
one can see a kink structure around $m_\chi \sim 400$~GeV,
which is close to the true LSP mass, $m_{\chi_1^0}=487$~GeV.
At the kink, the end point value of $m_{T2}$ is $m_{T2} \sim 1650$~GeV.
It should be noted that the inclusive $m_{T2}$ distribution is 
dominated by the events from the squark-gluino coproduction 
since the production cross section is larger than those of gluino-gluino
and squark-squark pair production.
In such a situation, the end point of $m_{T2}$ distributions
is sensitive to max$(m_{\tilde g},m_{\tilde q})$.
At point A, the gluino is heavier than the squarks and
the end point should be sensitive to the gluino mass, $m_{\tilde g}=1491$~GeV.
While the end point value is larger than 
the true gluino mass by about 150~GeV, 
we think the agreement between  $m_{T2}$ and $m^{(p)}_{T2}$  
is reasonable given the crudeness of our fit.

In the mSUGRA, the bino-like LSP mass is about 1/6 of the gluino mass. 
If we take the measured $m_{T2}$ end point at the kink as  the gluino mass 
then the LSP mass assuming mSUGRA is around  270~GeV. 
The observed kink is clearly above 270~GeV, therefore we can say 
that the mass spectrum is  the MMAM type for this case.  

\clearpage
\subsection{Monte Carlo analysis: Point B (mSUGRA)}
At  point B,  the squark is much heavier than the gluino and 
our interest is to measure the squark mass scale  
quantitatively using the inclusive $m_{T2}$ distributions. 
We require the following cuts to select the events.
\begin{enumerate}
 \item $n_{100}\ge 2$ 
and $n_{50}\ge 6$ 
within $|\eta|<3$. 
\item       $M_{\mathrm{eff}}>1500$~GeV
\item At least two jets in each hemisphere. 
\item $E_T^\mathrm{miss}>0.2 M_\mathrm{eff}$ and $E_T^\mathrm{miss}>
      100$~GeV.
\item There is no isolated lepton with $p_T>20$~GeV.
\end{enumerate}

\begin{figure}[htb]

    \centerline{
 \epsfxsize=\figscale\textwidth\epsfbox{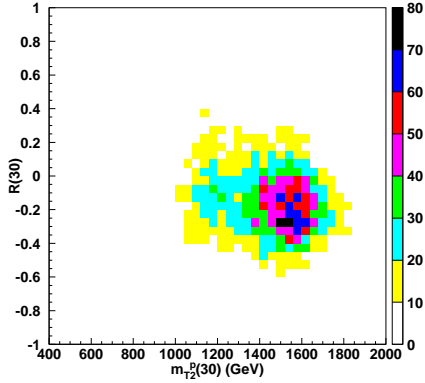}
}

    \caption{The two-dimensional distribution 
in $m_{T2}$-$R$ plane for point B. 
The test LSP mass is assumed as $m_{\chi}=30$~GeV.}
    \label{fig:ratio_sugra}
\end{figure}

In Fig.\ref{fig:ratio_sugra}, the two-dimensional distribution in 
the  $m^{(p)}_{T2}$-$R$ plane is plotted for $m_{\chi}=30$~GeV. 
The peak of the distribution appears $R\sim -0.2$, and
the misreconstruction rate is higher than point A. 
The low reconstruction efficiency  may be understood as follows.
At this point $m_{\tilde q} \sim 1850$~GeV,
$m_{\tilde g}\sim 1360$~GeV and 
the squark decaying into the gluino gives high $p_T$ jets as discussed earlier. 
When this jet is misidentified 
to the other hemisphere, the reconstructed $m_{T2}$ 
may become much lower than the expected $m^{(p)}_{T2}$
value, because the squark is so much heavier than 
gluino.  It  can be as low as the order of  gluino mass. Note that 
$(m_{\tilde q}-m_{\tilde g})/m_{\tilde q}\simeq 0.26$,  
roughly corresponds to the observed shift.   
Luckily, $m^{(p)}_{T2}$ strongly peaks near the end point, 
and the reconstructed event still makes visible end points.

\begin{figure}[htb]
    \centerline{
\epsfxsize=\figscale\textwidth\epsfbox{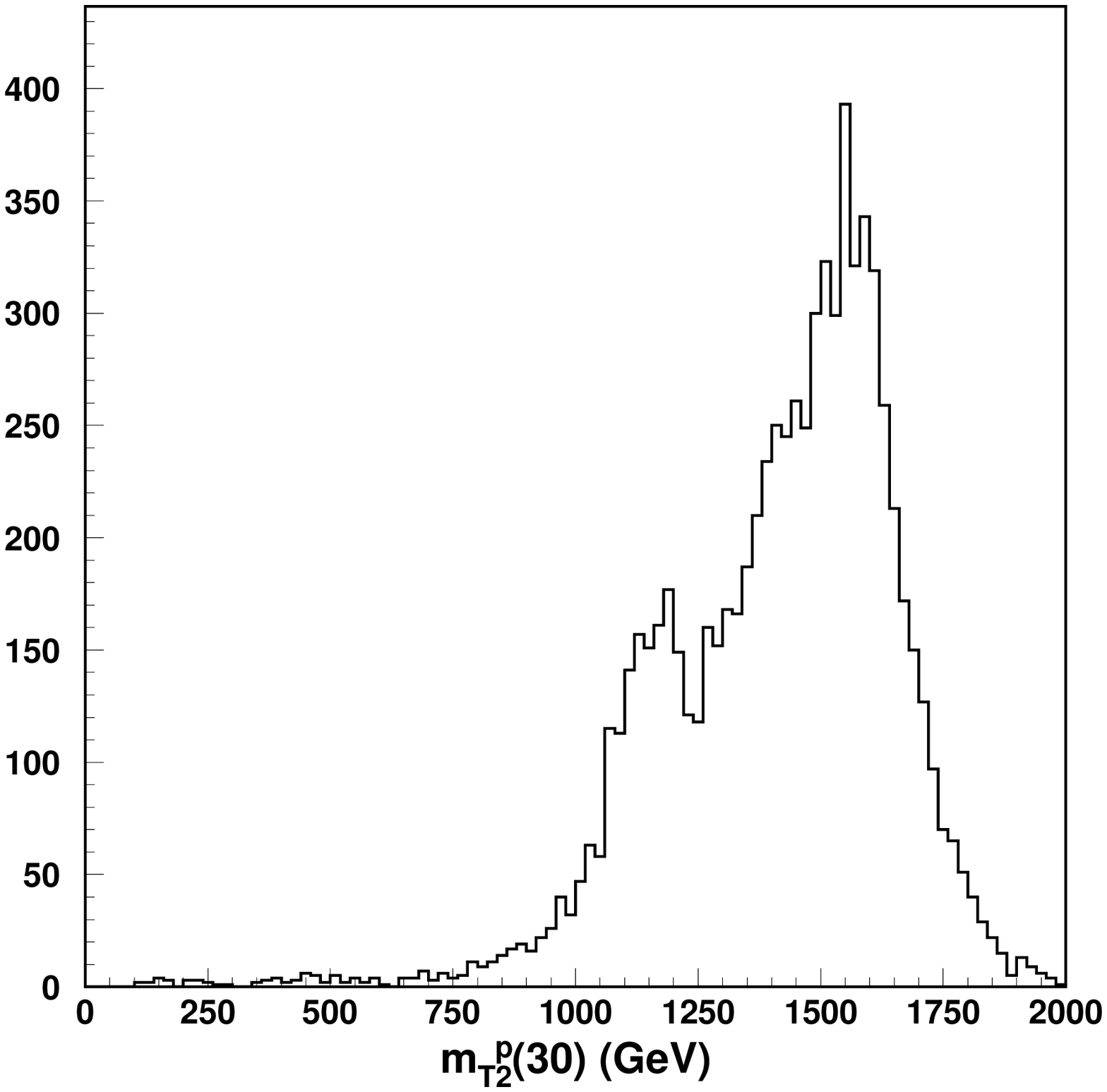}
\epsfxsize=\figscale\textwidth\epsfbox{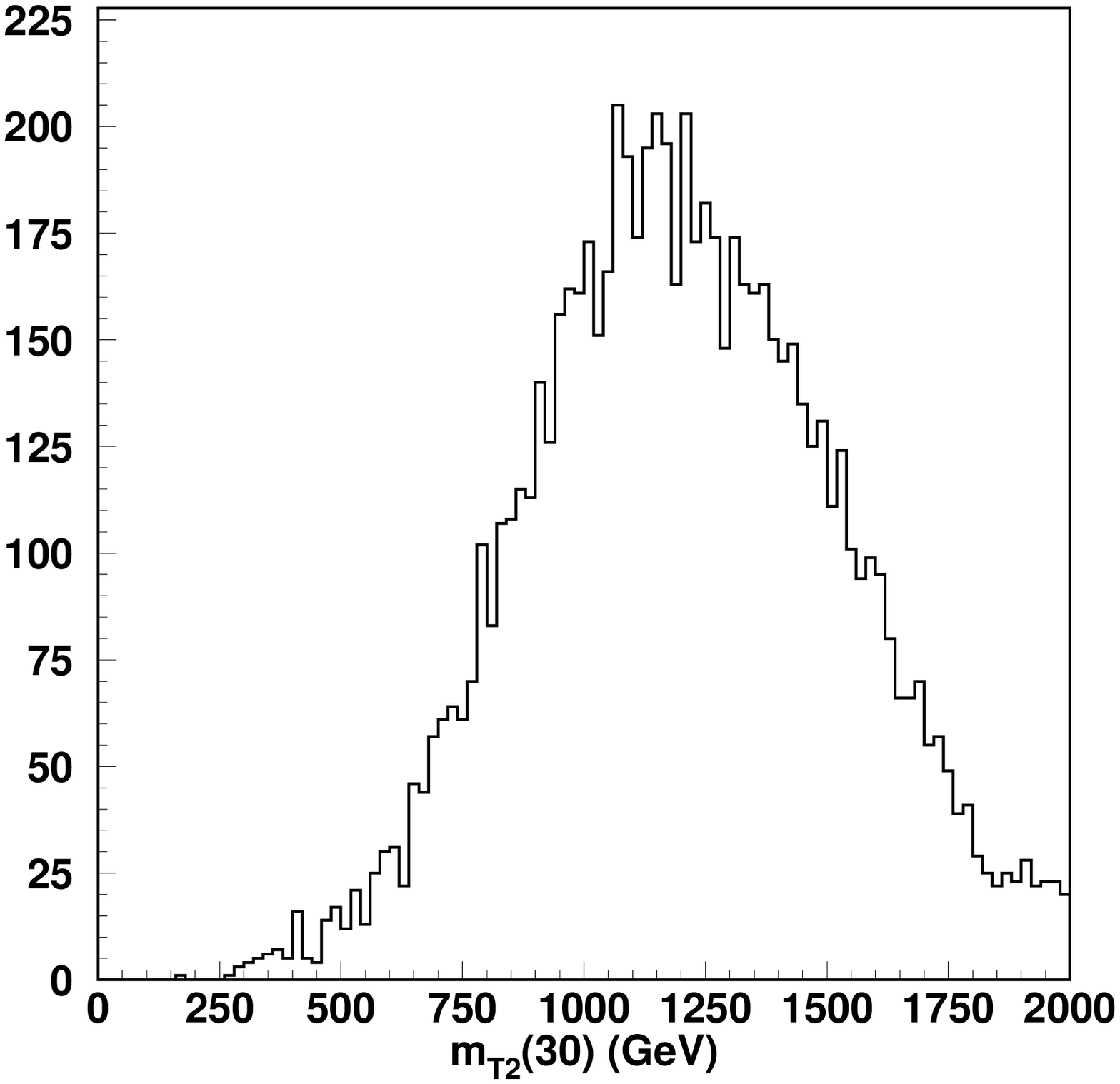}
}
    \caption{
(a). The $m^{(p)}_{T2}$ distribution for $m_\chi=30$~GeV.
(b). The  $m_{T2}$ distribution for $m_\chi=30$~GeV.
   }
    \label{fig:mt2_sugra}
\end{figure}

In Fig.\ref{fig:mt2_sugra}(a), the $m_{T2}^{(p)}$ distribution 
is plotted for $m_{\chi }=30$~GeV.
There are two peaks in the $m_{T2}^{(p)}$ distribution.
The higher peak corresponds to the squark while the lower peak
corresponds to the gluino. 
At point B, the squark is much heavier than the gluino, so the
end point is determined by the squark decay, $m_{T2}^{(p)}\sim 1850$~GeV.
In Fig.\ref{fig:mt2_sugra}(b), the $m_{T2}$ distribution
is shown for $m_\chi=30$~GeV.
The $m_{T2}$ distribution is smeared 
but  one can still see the end point. 
The end point events are dominated by the events with 
small $m^{vis}$. There is again interchange of the events near 
the end point as we increase the test LSP mass, and 
we can see the two peak structure in $m_{T2}(m_{\chi}> m_{\chi 1}^0)$ 
distribution for the event near the end point of 
 $m_{T2}(m_{\chi}< m_{\chi 1}^0)$  distribution, 
similar in Fig.\ref{fig:mt2_kklt_900a}.

\begin{figure}[htb]
    \centerline{{
\epsfxsize=0.45\textwidth\epsfbox{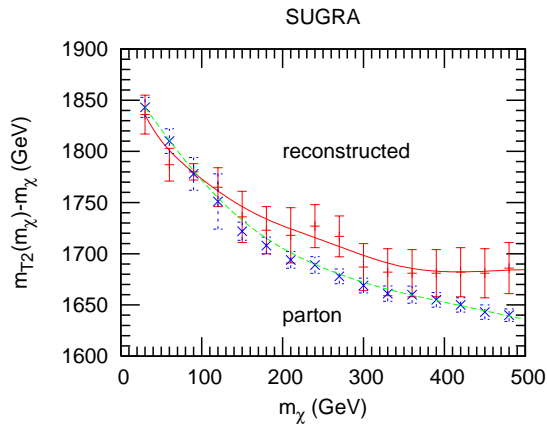}
}}
    \caption{The end point of $m_{T2}(m_\chi)-m_\chi$ 
 for various test LSP masses. The solid line is
 the $m_{T2}^\mathrm{max}$  while the  dashed line
 is the parton level $m_{T2}^{(p)\mathrm{max}}$. 
   }
    \label{fig:mt2max_sugra}
\end{figure}

\begin{figure}[htb]
    \centerline{{
\epsfxsize=0.45\textwidth\epsfbox{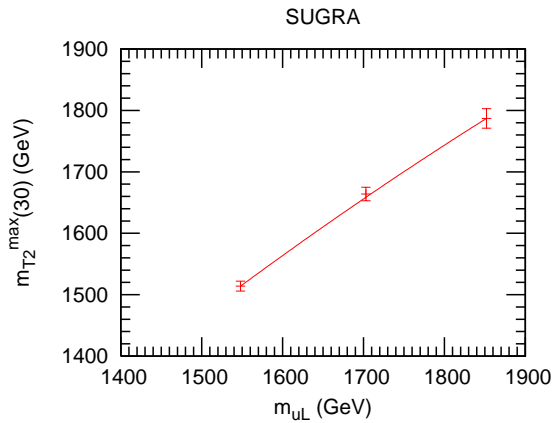}
}}
    \caption{ 
 The end point of $m_{T2}$ for $m_\chi=30$ at 
several mSUGRA points where the gaugino masses is the same 
as that of point B but $m_0$ is varied. 
Here the horizontal axis is the left-handed up-type squark mass.
   }
    \label{fig:msquark}
\end{figure}

In Fig.\ref{fig:mt2max_sugra}, the end points of $m_{T2}$ 
for various test LSP masses are plotted with a solid line. 
The end points are determined as in point A.
The end points of the $m_{T2}$ is almost the same
as the ones of $m_{T2}^{(p)}$ within errors.
One cannot  see a clear kink structure around the true LSP mass,
$m_{\chi_1^0}=237$~GeV. 
While it is difficult to determine the squark mass from the kink method,
the inclusive $m_{T2}$ analysis is useful to obtain the information on
the squark mass.
To see whether the end point of $m_{T2}$ correctly  describes the squark mass 
 for $m_{\tilde q}>m_{\tilde g}$, we show the $m_{T2}$ end point 
for $m_\chi=30$~GeV at 
the mSUGRA points where the gaugino mass is kept the same as 
that of point B but the universal scalar mass $m_0$ is varied.  
In Fig.\ref{fig:msquark}, we plot the $m_{T2}$ 
end point as the function of the squark mass and 
find very good agreement from 1400~GeV to 1800~GeV.

\clearpage 
\section{Summary and Conclusion}
\label{sec:sum}
In this paper, we have proposed an inclusive $m_{T2}$ analysis at the LHC 
to obtain information
on squark and gluino masses by the hemisphere method. 
The hemisphere method is  an algorithm to group collinear and high 
$p_T$ particles  and jets, assuming that there are two of such groups 
in a event. The algorithm is 
to group the cascade  decay products into two visible objects.
To study the distributions, 
 we perform the Monte Carlo simulation for  
 two sample SUSY spectra from the MMAM and the mSUGRA models.
The cascade decay products from a squark/gluino
are grouped into a visible object  with enough probability to see the 
parton level $m_{T2}$ end point. 
However, the end point of the $m_{T2}$ distribution 
is sometime smeared by mis-identification of hemispheres, which obscure
the end point determination.
We have fitted the $m_{T2}$ distribution near the end points.
The end point determination suffers from various systematic
uncertainties, such as a choice of the fitting function and the fitting region. 

We have examined the events near the end point  in detail.
For $m_\chi<m_{\chi_1^0}$ the end point events has the minimum
hemisphere mass,
$m^{\mathrm{vis}}_i=m^{\mathrm{vis}}_{\mathrm{min}}$.
The fake end point events due to
the mis-grouping of the hemisphere are reduced if we impose
the cut on the hemisphere mass, without disturbing the correct end points.
For $m_\chi>m_{\chi_1^0}$, 
 the end point event 
is realized when the hemisphere mass is at maximum 
$m^{\mathrm{vis}}_i=m^{\mathrm{vis}}_{\mathrm{max}}$.
The end point event is interchanged at $m_\chi=m_{\chi_1^0}$ and
the cusp structure of  $m_{T2}^{\mathrm{max}}$ appears.
By checking the test mass behavior of the $m_{T2}$ variable
for the events near the end  point,  we can prove if events near the
end point obtained by the fit of $m_{T2}$ distribution is 
correctly reconstructed ones or not. 
We have shown that the true end point event for $m_\chi<m_{\chi_1^0}$ 
gives the $m_{T2}$ value smaller than $m_{T2}^{max}$ for
$m_\chi\gg m_{\chi_1^0}$, while 
the fake end point event for $m_\chi<m_{\chi_1^0}$
gives $m_{T2}$ larger than $m_{T2}^{max}$ for
$m_\chi>m_{\chi_1^0}$.
From this observation, 
while there are various uncertainties for the end point determination,
we conclude that 
the inclusive $m_{T2}$ distribution is useful to obtain the information 
of the masses of the gluino/squark and the LSP 
at the LHC experiment.

For both of the sample points, the main  QCD production process of 
SUSY particles is 
squark-gluino coproduction, and the end point of $m_{T2}$ 
distribution is sensitive to max($m_{\tilde g}$, $m_{\tilde q}$).
For the sample point in the MMAM model, $m_{\tilde g}>m_{\tilde q}$
and the end point should be determined by $m_{\tilde g}$,
while it should be determined by $m_{\tilde q}$ 
for the sample point in the mSUGRA model because 
$m_{\tilde q}>m_{\tilde g}$.
From the Monte Carlo analysis, we have found that the end point
is indeed determined by max($m_{\tilde g}$, $m_{\tilde q}$)
for both of the sample points.
For the MMAM sample point,
we have found that there is a cusp-like structure of 
$m_{T2}^{\mathrm {max}}(m_{\chi})$ at the true LSP mass and we can
determine the gluino and the LSP mass simultaneously.
For the mSUGRA sample points with $m_{\tilde q}>m_{\tilde g}$, we find 
the linearity between $m_{\tilde{q}}$ and $m_{T2}^{\rm max}$.  
We have checked that  the squark mass 
is reconstructed 
up to $m_{\tilde{q}} \sim1.4m_{\tilde{g}}$ when $m_{\tilde{g}}\sim 1.4$~TeV.  

There have been different approaches in the LHC physics study. 
One of the direction is to study the inclusive quantities such as 
$M_{\rm eff}$, $E^{\rm miss }_{T}$, which do not require reconstruction 
and is useful to grab  characters of the events. The other direction 
is to study quantities, which are specific to some processes,
such as the end point measurements
 of the invariant mass and the $m_{T2}$  distribution.
They are very powerful to determine the absolute sparticle masses.   
In this paper, we propose an inclusive $m_{T2}$, which is inclusive 
in the sense that we do not specify the decay channel.  
However, it  bites the merit of exclusive 
analyses with  helps of the new 
understandings  of  the stransverse mass function $m_{T2}(m_{\chi})$ .  
While  detailed analyses 
on  systematical uncertainties are still needed,   we hope 
that this quantity helps to determine the sparticle masses 
at the early stage of the LHC experiments.

\section*{Acknowledgement}

This work is supported in part by the Grant-in-Aid for Science Research,
Ministry of Education, Culture, Sports, Science and Technology,
Japan (No.16081207, 18340060 for M.M.N.).

\newpage
%
%
\newcommand{\Journal}[4]{{\sl #1} {\bf #2} {(#3)} {#4}}
\newcommand{\APJ}{Ap. J.}
\newcommand{\CJP}{Can. J. Phys.}
\newcommand{\MPL}{Mod. Phys. Lett.}
\newcommand{\NC}{Nuovo Cimento}
\newcommand{\NP}{Nucl. Phys.}
\newcommand{\PL}{Phys. Lett.}
\newcommand{\PR}{Phys. Rev.}
\newcommand{\PRep}{Phys. Rep.}
\newcommand{\PRL}{Phys. Rev. Lett.}
\newcommand{\PTP}{Prog. Theor. Phys.}
\newcommand{\SJNP}{Sov. J. Nucl. Phys.}
\newcommand{\ZP}{Z. Phys.}

\end{document}